\documentclass[aip,
amsmath,amssymb,numerical,
preprint,%
]{revtex4-1}
\usepackage{graphicx}
\usepackage{dcolumn}
\usepackage{bm}
\usepackage{hyperref}
\usepackage[utf8]{inputenc}
\usepackage[T1]{fontenc}
\usepackage{mathptmx}
\usepackage{etoolbox}
\usepackage{array,xcolor,colortbl}
\usepackage{multirow}
\makeatletter
\def\@email#1#2{%
	\endgroup
	\patchcmd{\titleblock@produce}
	{\frontmatter@RRAPformat}
	{\frontmatter@RRAPformat{\produce@RRAP{*#1\href{mailto:#2}{#2}}}\frontmatter@RRAPformat}
	{}{}
}%

\makeatother

\begin{document}
	
	\title{Phototactic bioconvection with the effect of oblique collimated flux at forward scattering algae suspension in rotating medium}
	\author{S. K. Rajput}
	\altaffiliation[Corresponding author email: ]{shubh.iitj@gmail.com}
	\affiliation{ 
		Department of Mathematics, PDPM Indian Institute of Information Technology Design and Manufacturing,
		Jabalpur 482005, India.
	}%
	
	
	
	
	\begin{abstract}
		
		The primary objective of this article is to explore how rotation influences the initiation of phototactic bioconvection. This investigation is conducted through the application of linear stability theory to a suspension composed of forward-scattering phototactic algae. The suspension is uniformly exposed to oblique collimated flux. The bioconvection phenomenon is characterized by an unstable disturbance mode that undergoes a transition from a stationary state to an oscillatory state as the Taylor number varies while keeping other parameters constant. Additionally, it is noteworthy that rotation of the system has a substantial stabilizing effect on the suspension.
	\end{abstract}
	
	
	\maketitle
	
	
	\section{INTRODUCTION}
	
	Bioconvection, a well-recognized occurrence in fluid dynamics, pertains to the development of patterns in suspensions containing living microorganisms like algae and bacteria~\cite{20platt1961,21pedley1992,22hill2005,23bees2020,24javadi2020}. Platt~\cite{20platt1961} initially coined the term "bioconvection" in 1961. Typically, these microorganisms, albeit on a small scale, have a greater density than the surrounding fluid and collectively propel themselves upward. Nevertheless, in certain natural settings, density disparities aren't mandatory for pattern formation, and microorganisms exhibit varied behaviours. Inanimate microorganisms do not display this pattern-forming behaviour. Microorganisms react to external stimuli, leading to changes in their swimming direction, known as taxis. Examples of taxis include gravitaxis, which is controlled by gravity, gyrotaxis, which depends on gravity and viscous torque when bottom-heavy microorganisms are present, and phototaxis, which is the movement in the direction of or away from light sources. The effects of phototaxis are a specific emphasis of this article.
	
	Significant research efforts have been devoted to investigating the occurrence of bioconvection patterns in algal suspensions, with a particular focus on the profound influence of light on fluid dynamics and cell distribution~\cite{1wager1911,2kitsunezaki2007}. Dynamic patterns can either survive or be broken in well-mixed cultures of photosynthetic microorganisms when exposed to intense light. Intense light has the power to prevent the production of stable patterns and destroy ones that have already formed~\cite{3kessler1985,4williams2011,5kessler1989}. A key factor in modifying bioconvection patterns is how algae react to different light levels. Furthermore, activities like light scattering and absorption are a part of the interaction between light and microorganisms. Depending on cell properties including size, shape, and refractive index, algal light scattering can be isotropic or anisotropic, with the latter further subdivided into forward and backward scattering. Algae mostly display forward scattering of light in the visible spectrum due to their size.

	Rajput's phototaxis model is used in the study to examine the bioconvection system. In this study, a suspension of phototactic microorganisms is assumed to rotate about the z-axis at a constant angular velocity while being irradiated from above by an oblique collimated flux. Numerous motile algae rely on photosynthesis to meet their energy needs, which causes them to behave in a certain way called phototaxis. It is crucial to look at the phototaxis model in the context of a rotating medium in order to fully comprehend how these microorganisms react to light. Our method includes anisotropic scattering, in contrast to earlier models that viewed scattering effects as isotropic. With this alteration, the movement of microorganisms in response to light stimuli may be more precisely shown. Our aim is to improve the model's accuracy and give useful insights into the complex dynamics of the bioconvection phenomena by taking the effect of anisotropic scattering into account.
	
	\begin{figure}[!ht]
		\centering
		\includegraphics[width=14cm]{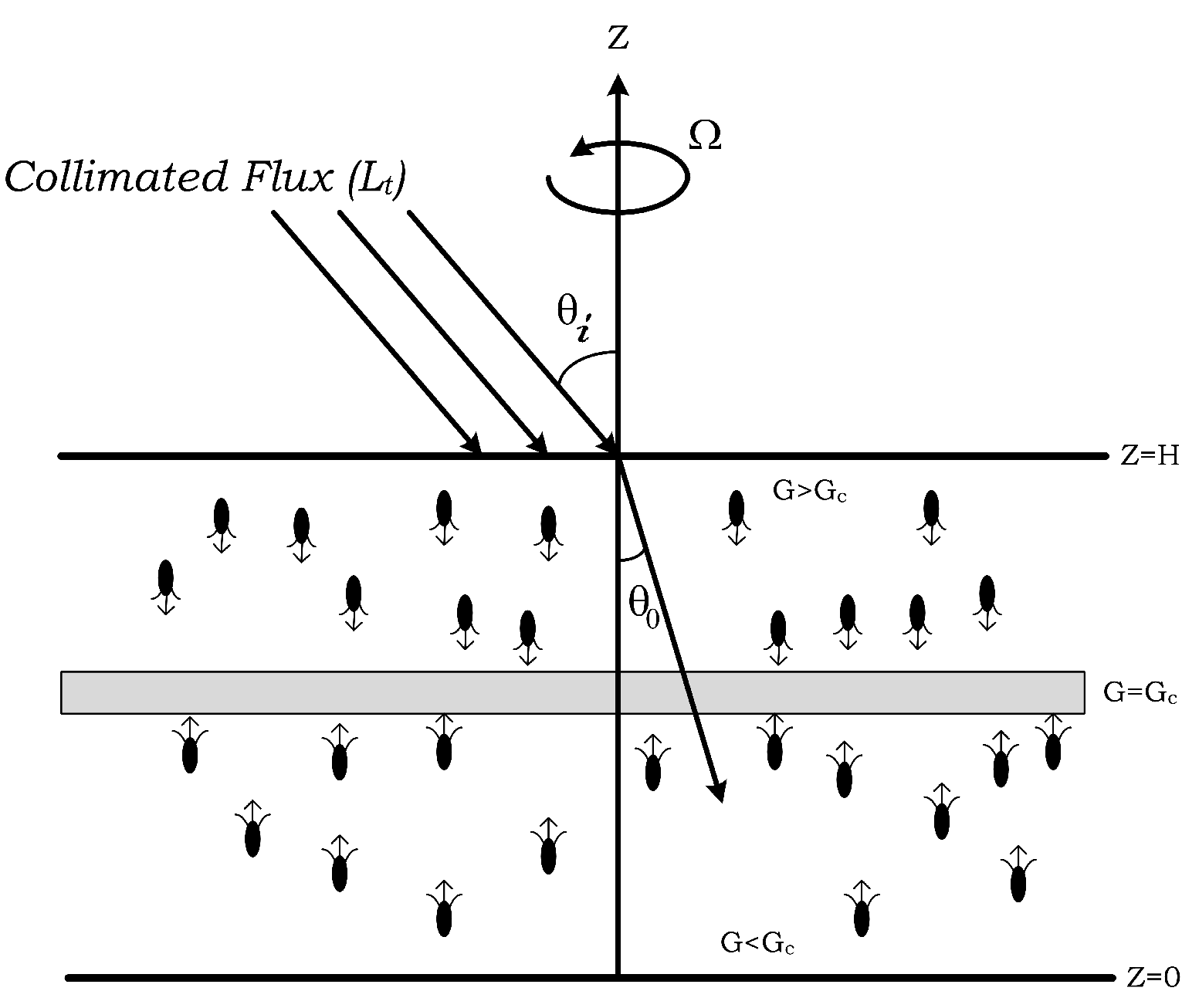}
		\caption{\footnotesize{Formation of sublayer in a rotating medium.}}
		\label{fig1}
	\end{figure}
	
	Between two parallel boundaries at $z=0$ and $z=H$, there is a suspension of algae that mostly scatters light in the forward direction. Oblique collimated light is used to uniformly illuminate this suspension from above. When a steady-state condition is reached, which shows no net fluid movement, the system is at its equilibrium state. In this equilibrium condition, the horizontal sublayer that forms as a result of a balance between phototaxis (movement in reaction to light) and diffusion results in the accumulation of cells, essentially splitting the suspension into two different zones. The location of this sublayer is greatly influenced by the critical intensity, represented as $G_c$. Where the light intensity is more than $G_c$ ($G > G_c$) above the sublayer, the light is strong enough to prevent cell migration. In contrast, cells below the sublayer ($G< G_c$) move upward in the direction of the sublayer as a result of their phototactic reaction. The sublayer's lowest portion is thought to be unstable and fluid motion is possible there. This fluid motion may cause penetrative convection, a phenomenon seen in a variety of convection issues if the system becomes unstable. Penetrative convection occurs when an unstable lower zone penetrates a higher, stable region, perhaps causing mixing and circulation.
	
	Over time, researchers have embarked on a fascinating journey to unravel the mysteries of phototactic bioconvection. For instance, Vincent and Hill~\cite{12vincent1996} took on the challenge of deciphering how phototactic bioconvection begins. They performed a linear stability analysis and found oscillatory and non-oscillatory behaviour of the solution. Meanwhile, Ghorai and Hill~\cite{10ghorai2005} decided to reveal the cause of bioconvective flow patterns in two dimensions. They used Vincent and Hill's model with a non-linear response of algae cells to light and found the same results. Ghorai $et$ $al$.~\cite{7ghorai2010} explored the genesis of bioconvection in isotropic scattering suspensions. What they uncovered was truly mesmerizing—an ever-shifting equilibrium marked by a sublayer residing at different depths, a phenomenon shaped by isotropic scattering. As the journey continued, Panda and Ghorai~\cite{14panda2013} discovered the linear stability within forward-scattering algal suspensions. Panda and Singh~\cite{11panda2016} conducted numerical explorations of linear stability using Vincent and Hill's continuum model within a two-dimensional realm.  Panda $et$ $al$.~\cite{15panda2016} ventured into the realm of diffuse radiation, illuminating its profound impact on isotropic scattering algal suspensions. Their findings revealed how this diffuse illumination became a stabilizing force. Panda~\cite{8panda2020} continued to probe deeper. This researcher navigated the realms of forward anisotropic scattering, not only considering diffuse and collimated irradiation but also unearthing the profound implications of this interplay on the onset of phototactic bioconvection. Panda $et$ $al$.~\cite{16panda2022} introduced a new element—an oblique collimated flux—that sparked a transformation in the bioconvective solutions. They uncovered a mesmerizing dance between non-oscillatory and overstable states, a phenomenon brought on by the influence of oblique collimated flux. After that, Kumar~\cite{17kumar2022} set to explore the effects of oblique collimated irradiation on isotropic scattering algal suspensions. He discovered both stable and overstable solutions. After that, Panda and Rajput~\cite{41rajput2023} investigated the impact of both diffuse and oblique collimated fluxes. Kumar~\cite{40kumar2023} investigated the impact of collimated flux at the onset of bioconvection in the rotating medium and uncovered—a stabilizing effect due to rotation in the phototactic bioconvection, even within a non-scattering medium. Rajput forwarded his research in an isotropic and forward scattering medium. After that, Rajput developed models in which he investigated the impact of oblique collimated flux in the rotating non-scattering and scattering (isotropic) medium. The impact of oblique collimated flux in a rotating forward scattering medium has not been investigated till now. Therefore, in this article, the impact of oblique collimated flux in a rotating forward scattering medium is investigated for the deep knowledge of phototactic bioconvection in different conditions.
	
	The article adopts a methodical approach, commencing with the mathematical formulation of the problem at hand. It meticulously derives the equilibrium solution and then introduces small disturbances into the governing system of bioconvection. These perturbations lay the foundation for a comprehensive exploration of the linear stability problem, which is subsequently tackled and resolved through the application of numerical methods. The results obtained from this rigorous analysis are meticulously presented and subjected to thorough discussion.
	
	
	\section{MATHEMATICAL FORMULATION}
	
	The article is primarily concerned with the examination of a forward-scattering algal suspension's behaviour within the (y, z) plane. This suspension is confined between two parallel horizontal boundaries. Importantly, these boundaries are configured in a way that there is no reflection of light occurring from either the top or bottom boundaries. This specific configuration allows for a thorough investigation into how the suspension behaves when exposed to light. In this setup, the suspension experiences uniform illumination from above through oblique collimated flux. 
	
	\begin{figure}[!ht]
		\centering
		\includegraphics[width=14cm]{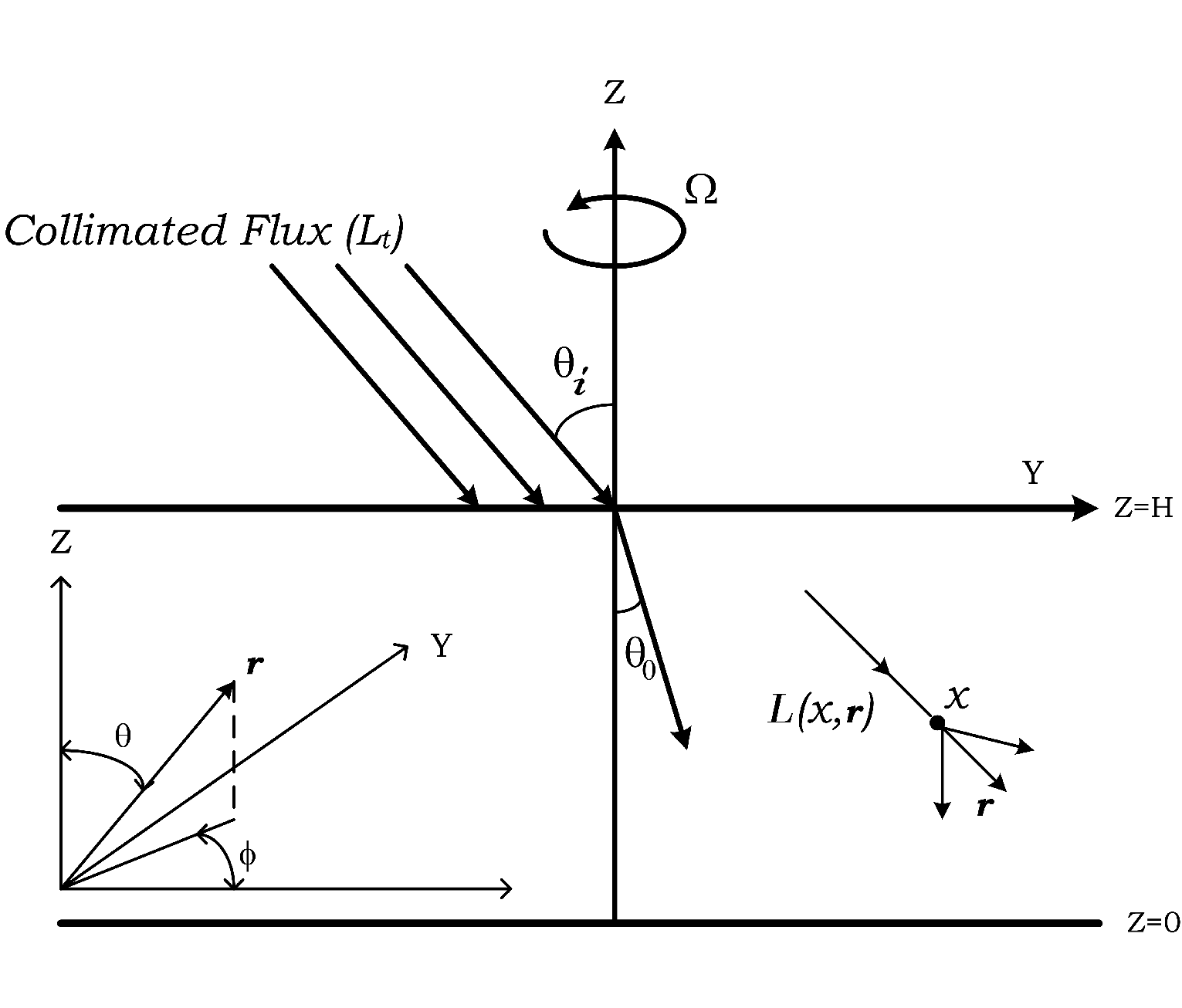}
		\caption{\footnotesize{Formation of sublayer in a rotating medium.}}
		\label{fig2}
	\end{figure}
	
	
	\subsection{THE AVERAGE SWIMMING DIRECTION}
	
	The Radiative Transfer Equation (RTE) serves as a mathematical tool to describe how light behaves and interacts within the medium under consideration. It can be represented as follows
	
	\begin{equation}\label{1}
		\frac{dL(\boldsymbol{x},\boldsymbol{r})}{dr}+(a+\sigma_s)L(\boldsymbol{x},\boldsymbol{r})=\frac{\sigma_s}{4\pi}\int_{0}^{4\pi}L(\boldsymbol{x},\boldsymbol{r'})\Xi(\boldsymbol{r},\boldsymbol{r'})d\Omega',
	\end{equation}
	
	in which $a$ and $\sigma_s$ correspond to the absorption coefficient and scattering coefficient, respectively. The scattering phase function denoted as $\Xi(\boldsymbol{r},\boldsymbol{r'})$, characterizes how light spreads in different directions after being scattered from $\boldsymbol{r'}$ to $\boldsymbol{r}$ direction. We assume that the scattering phase function displays linear anisotropy with azimuthal symmetry for the sake of this investigation. In more precise terms, $\Xi(\boldsymbol{r},\boldsymbol{r'})$ is written as $\Xi(\boldsymbol{r},\boldsymbol{r'})=1+A\cos\theta\cos\theta'$. The anisotropy's degree in this case is determined by the parameter $A$, which acts as the anisotropy coefficient. Forward scattering is indicated by positive values of $A$, backward scattering by negative values, and isotropic scattering by $A=0$.
	
	The light intensity at the upper surface of the suspension is described as follows
	
	\begin{equation*}
		L(\boldsymbol{x}_b,\boldsymbol{r}) = L_t\delta(\boldsymbol{r}-\boldsymbol{r}_0).
	\end{equation*}
	
	Here, an arbitrary location on the upper boundary surface is indicated by the $\boldsymbol{x}_b=(x,y, H)$, and the $\boldsymbol{ r}_0=\sin(\pi-\theta_0)[\cos\phi+\sin\phi]+\cos(\pi-\theta_0)$. The oblique collimated flux magnitude is represented by $L_t$, the zenith angle is represented by $\theta$, and the zenith angle value is represented by $\theta_0$.
	
	Now two variables have been considered, $a=\alpha n(\boldsymbol{x})$ and $\sigma_s=\beta n(\boldsymbol{x})$, the Radiative Transfer Equation (RTE) may be expressed as follows
	
	\begin{equation}\label{2}
		\frac{dL(\boldsymbol{x},\boldsymbol{r})}{dr}+(\alpha+\beta)nL(\boldsymbol{x},\boldsymbol{r})=\frac{\beta n}{4\pi}\int_{0}^{4\pi}L(\boldsymbol{x},\boldsymbol{r'})(A\cos{\theta}\cos{\theta'})d\Omega'.
	\end{equation}
	
	The total intensity at a specific point $\boldsymbol{x}$ within the medium can be expressed as
	
	\begin{equation*}
		G(\boldsymbol{x}) = \int_0^{4\pi}L(\boldsymbol{x},\boldsymbol{r})d\Omega.
	\end{equation*}
	
	Similarly, the radiative heat flux is defined as
	\begin{equation}\label{3}
		\boldsymbol{q}(\boldsymbol{x}) = \int_0^{4\pi}L(\boldsymbol{x},\boldsymbol{r})\boldsymbol{r}d\Omega.
	\end{equation}
	
	The mean swimming velocity of microorganisms is determined as
	
	\begin{equation*}
		\boldsymbol{W}_c = W_c<\boldsymbol{p}>.
	\end{equation*}
	
	The calculation of the mean swimming orientation of the cell, denoted as $<\boldsymbol{p}>$, involves the use of the following equation
	
	\begin{equation}\label{4}
		<\boldsymbol{p}> = -M(G)\frac{\boldsymbol{q}}{|\boldsymbol{q}|},
	\end{equation}
	
	in which $M(G)$ is the taxis function, which describes how algal cells react to light in this case. It's described as
	
	\begin{equation*}
		M(G)=\left\{\begin{array}{ll}\geq 0, & \mbox{ } G(\boldsymbol{x})\leq G_{c}, \\
			< 0, & \mbox{ }G(\boldsymbol{x})>G_{c}.  \end{array}\right. 
	\end{equation*}
	
	The taxis function plays a crucial role in determining how cells respond to changes in light intensity. It governs the behaviour of the cells in response to variations in light intensity.
	
	\subsection{GOVERNING EQUATIONS WITH BOUNDARY CONDITIONS}
	
	The system is controlled by a number of equations in the model of a suspension containing phototactic microorganisms that are provided. The following are these equations and the corresponding boundary conditions:
	
	Mass conservation equation:
	
	\begin{equation}\label{5}
		\boldsymbol{\nabla}\cdot \boldsymbol{u}=0,
	\end{equation}
	
	momentum equation under the Boussinesq approximation
	
	\begin{equation}\label{6}
		\rho\left(\frac{\partial \boldsymbol{u}}{\partial t}+(\boldsymbol{u}\cdot\nabla )\boldsymbol{u}+2\boldsymbol{\Omega}\times \boldsymbol{u}\right)=-\boldsymbol{\nabla} P_e+\mu\nabla^2\boldsymbol{u}-nvg\Delta\rho\hat{\boldsymbol{z}}m
	\end{equation}
	
	cell conservation equation
	
	\begin{equation}\label{7}
		\frac{\partial n}{\partial t}=-\boldsymbol{\nabla}\cdot \boldsymbol{B},
	\end{equation}
	
	total cell flux equation
	
	\begin{equation}\label{8}
		\boldsymbol{B}=n\boldsymbol{u}+nW_c<\boldsymbol{p}>-\boldsymbol{D}\boldsymbol{\nabla} n,
	\end{equation}
	
	where $\boldsymbol{u}$ denotes the average fluid speed, $n$ represents the concentration of algal cells per unit volume, $v$ signifies the cellular volume, $\rho$ stands for the water density, $\Delta\rho/\rho$ expresses the slight variation in cell density, $\boldsymbol{\Omega}$ is the angular velocity, $\mu$ represents the dynamic viscosity of the fluid mixture, and $\boldsymbol{D}=DI$ indicates the diffusivity of the cells.
	
	The boundary conditions for the suspension are as follows
	
	\begin{equation}\label{9}
		\boldsymbol{u}\cdot\hat{\boldsymbol{z}}= \boldsymbol{u}\times\hat{\boldsymbol{z}}=\boldsymbol{B}\cdot\hat{\boldsymbol{z}}=0\quad \text{at}\quad z=0,
	\end{equation}
	\begin{equation}\label{10}
		\boldsymbol{u}\cdot\hat{\boldsymbol{z}}=\frac{\partial^2}{\partial z^2}(\boldsymbol{u}\cdot\hat{\boldsymbol{z}})=\boldsymbol{B}\cdot\hat{\boldsymbol{z}}=0\quad \text{at}\quad z=H.
	\end{equation}
	
	Furthermore, the upper boundary experiences an oblique collimated flux, resulting in distinct conditions for the intensity at the boundaries
	
	\begin{subequations}
		\begin{equation}\label{11a}
			L(x, y, z = 0, \theta, \phi)=L_t\delta(\boldsymbol{r}-\boldsymbol{r_0}),\quad \left(\frac{\pi}{2}\leq\theta\leq\pi\right),
		\end{equation}
		\begin{equation}\label{11b}
			L(x, y, z = 0, \theta, \phi) =0,\quad \left(0\leq\theta\leq\frac{\pi}{2}\right).
		\end{equation}
	\end{subequations}
	
	We establish the following reference scales: length scale $H$, time scale $H^2/D$, velocity scale $D/H$, pressure scale $\mu D/H^2$, and concentration scale $\bar{n}$ in order to express the governing equations in a dimensionless way. We can now move on to non-dimensionalize the governing equations with these scales
	
	\begin{equation}\label{12}
		\boldsymbol{\nabla}\cdot\boldsymbol{u}=0,
	\end{equation}
	\begin{equation}\label{13}
		Sc^{-1}\left(\frac{\partial \boldsymbol{u}}{\partial t}+(\boldsymbol{u}\cdot\nabla )\boldsymbol{u}\right)+\sqrt{Ta}(\hat{z}\times\boldsymbol{u})=-\nabla P_{e}-Rn\hat{\boldsymbol{z}}+\nabla^{2}\boldsymbol{u},
	\end{equation}
	\begin{equation}\label{14}
		\frac{\partial{n}}{\partial{t}}=-\boldsymbol{\nabla}\cdot\boldsymbol{B}=-{\boldsymbol{\nabla}}\cdot[\boldsymbol{n{\boldsymbol{u}}+nV_{c}<{\boldsymbol{p}}>-{\boldsymbol{\nabla}}n.}]
	\end{equation}
	
	The Rayleigh number $R$ is defined as $R=\bar{n}vg\Delta\rho H^3/\nu\rho D$, the Schmidt number $Sc=\nu/D$, and $Ta=4\Omega^2H^4/\nu^2$ is the Taylor number. The boundary conditions can be represented in a dimensionless form after the non-dimensionalization procedure as follows
	
	\begin{equation}\label{15}
		\boldsymbol{u}\cdot\hat{z}=\boldsymbol{u}\times\hat{z}=B\cdot\hat{z}=0 \quad \text{at} \quad z=0,
	\end{equation}
	\begin{equation}\label{16}
		\boldsymbol{u}\cdot\hat{z}=\frac{\partial^2}{\partial z^2}(\boldsymbol{u}\cdot\hat{z})=B\cdot\hat{z}=0 \quad \text{at} \quad z=1.
	\end{equation}
	
	It is possible to define the dimensionless Radiative Transfer Equation (RTE) as
	
	\begin{equation}\label{17}
		\frac{dL(\boldsymbol{x},\boldsymbol{r})}{dr}+\kappa nL(\boldsymbol{x},\boldsymbol{r})=\frac{\sigma n}{4\pi}\int_{0}^{4\pi}L(\boldsymbol{x},\boldsymbol{r'})(A\cos{\theta}\cos{\theta'})d\Omega',
	\end{equation}
	
	in which the dimensionless extinction and scattering coefficients are $\kappa=(\alpha+\beta)\bar{n}H$ and $\sigma=\beta\bar{n}H$, respectively. The scattering albedo $\omega=\sigma/(\kappa+\sigma)$ measures how well microorganisms scatter light. Alternatively, direction cosines may be used to represent the Radiative Transfer Equation (RTE), which provides a practical way to describe the direction of light propagation
	
	\begin{equation}\label{18}
		\xi\frac{dL}{dx}+\eta\frac{dL}{dy}+\nu\frac{dL}{dz}+\kappa nL(\boldsymbol{x},\boldsymbol{r})=\frac{\omega\kappa n}{4\pi}\int_{0}^{4\pi}L(\boldsymbol{x},\boldsymbol{r'})(A\cos{\theta}\cos{\theta'})d\Omega'.
	\end{equation}
	
	The intensity at the boundary is denoted by the following in the dimensionless formulation
	
	\begin{subequations}
		\begin{equation}\label{19a}
			L(x, y, z = 1, \theta, \phi)=L_t\delta(\boldsymbol{r}-\boldsymbol{r_0}),\qquad \left(\frac{\pi}{2}\leq\theta\leq\pi\right),
		\end{equation}
		\begin{equation}\label{19b}
			L(x, y, z = 0, \theta, \phi) =0,\qquad \left(0\leq\theta\leq\frac{\pi}{2}\right).
		\end{equation}
	\end{subequations}
	
	
	\section{THE STEADY STATE SOLUTION}
	
	The steady state of the system is defined as
	
	\begin{equation*}
		\boldsymbol{u}=0, \quad n=n_b(z), \quad \zeta_b=\nabla\times \boldsymbol{u}=0, \quad \text{and} \quad L=L_b(z,\theta).
	\end{equation*}
	
	The total intensity $G_b$ and radiative flux $\boldsymbol{q_b}$ may be represented as follows at equilibrium
	
	\begin{equation*}
		G_b=\int_0^{4\pi}L_b(z,\theta)d\Omega, \quad \text{and} \quad \boldsymbol{q}_s=\int_0^{4\pi}L_b(z,\theta)\boldsymbol{s}d\Omega.
	\end{equation*}
	
	Due to $L_b(z,\theta)$'s independence from $\phi$, the vector $\boldsymbol{q_b}$ has zero components in the x and y directions. As a result, $\boldsymbol{q_b}=-q_b\hat{\boldsymbol{z}}$ may be used to represent $\boldsymbol{q_b}$, where $q_b$ stands for the magnitude of $\boldsymbol{q_b}$.
	
	Given by is the equation regulating $L_b$ at equilibrium
	
	\begin{equation}\label{20}
		\frac{dL_b}{dz}+\frac{\kappa n_bL_b}{\nu}=\frac{\omega\kappa n_b}{4\pi\nu}(G_b(z)-Aq_s\nu).
	\end{equation}
	
	The collimated part of the equilibrium intensity $L_b^c$ and the diffuse part $L_b^d$ brought on by scattering may be distinguished from one another. The governing equation for the collimated component, $L_b^c$, is
	
	\begin{equation}\label{21}
		\frac{dL_b^c}{dz}+\frac{\kappa n_bL_b^c}{\nu}=0,
	\end{equation}
	
	with the boundary condition
	
	\begin{equation}\label{22}
		L_b^c(1, \theta) =L_t\delta(\boldsymbol{r}-\boldsymbol{r}_0), \quad \theta\in[\pi/2,\pi].
	\end{equation}
	
	The expression for $L_b^c$ is obtained by solving the governing equation for $L_b^c$ with the given boundary condition
	
	\begin{equation}\label{23}
		L_b^c=L_t\exp\left(\int_z^1\frac{\kappa n_b(z')}{\nu}dz'\right)\delta(\boldsymbol{r}-\boldsymbol{r}_0).
	\end{equation}
	
	Equation governs the diffuse component $L_b^d$
	
	\begin{equation}\label{24}
		\frac{dL_b^d}{dz}+\frac{\kappa n_bL_b^d}{\nu}=\frac{\omega\kappa n_b}{4\pi\nu}(G_b(z)-Aq_s\nu),
	\end{equation}
	
	with the boundary conditions
	
	\begin{subequations}
		\begin{equation}\label{25a}
			L_b^d(1, \theta) =0, \quad \theta\in[\pi/2,\pi],
		\end{equation}
		\begin{equation}\label{25b}
			L_b^d(0, \theta) =0, \quad \theta\in[0,\pi/2].
		\end{equation}
	\end{subequations}
	
	The overall intensity $G_b$ in the initial state is calculated as
	
	\begin{equation}\label{26}
		G_b=G_b^c+G_b^d=\int_0^{4\pi}[L_b^c(z,\theta)+L_b^d(z,\theta)d]\Omega=L_t\exp\left(\frac{-\int_z^1\kappa n_b(z')dz'}{\cos\theta_0}\right)+\int_0^{\pi}L_b^d(z,\theta)d\Omega.
	\end{equation}
	
	In the fundamental state, the collimated and diffuse components of the total intensity $G_b$ are denoted by $G_b^c$ and $G_b^d$, respectively.
	
	Similar to this, the radiative heat flux in the basic state may be expressed as
	
	\begin{equation}\label{27}
		\boldsymbol{q_b}=\boldsymbol{q_b}^c+\boldsymbol{q_b}^d=\int_0^{4\pi}\left(L_b^c(z,\theta)+L_b^d(z,\theta)\right)\boldsymbol{r}d\Omega=-L_t(\cos\theta_0)\exp\left(\frac{\int_z^1-\kappa n_b(z')dz'}{\cos\theta_0}\right)\hat{\boldsymbol{z}}+\int_0^{4\pi}L_b^d(z,\theta)\boldsymbol{r}d\Omega.
	\end{equation}
	
	In order to solve this problem, a new variable called $\tau$ is established
	
	\begin{equation*}
		\tau=\int_z^1 \kappa n_b(z')dz'.
	\end{equation*}
	
	These equations make up a collection of linked second-order Fredholm integral equations. The equation for $G_b$ and $q_b$ may be written as follows
	
	\begin{equation}\label{28}
		G_b(\tau) = \frac{\omega}{2}\int_0^{\kappa} G_b(\tau')E_1(|\tau-\tau'|)d\tau'+e^{-\tau/\cos\theta_0}+A sgn(\tau-\tau')q_b(\tau')E_2(|\tau-\tau'|),
	\end{equation}
	
	\begin{equation}\label{29}
		q_b(\tau) = \frac{\omega}{2}\int_0^{\kappa} Aq_b(\tau')E_3(|\tau-\tau'|)d\tau'+(\cos\theta_0)e^{-\tau/\cos\theta_0}+sgn(\tau-\tau')G_b(\tau')E_2(|\tau-\tau'|),
	\end{equation}
	
	where $sgn(x)$ is the signum function and $E_n(x)$ is the exponential integral of order $n$. One way for resolving these coupled Fredholm integral equations is the process of singularity removal.
	
	The fundamental state's average swimming direction is provided by
	
	\begin{equation*}
		<\boldsymbol{p_b}>=-M_b\frac{\boldsymbol{q_b}}{q_b}=M_b\hat{\boldsymbol{z}},
	\end{equation*}
	
	in which $M_b=M(G_b)$. The following equation is satisfied by the cell concentration in the equilibrium state
	
	\begin{equation}\label{30}
		\frac{dn_b}{dz}-V_cM_bn_b=0,
	\end{equation}
	
	accompanied by the equation
	
	\begin{equation}\label{31}
		\int_0^1n_b(z)dz=1.
	\end{equation}
	
	Equations (\ref{28}) through (\ref{31}) together constitute a boundary value problem that may be resolved numerically using techniques such as the shooting method.
	
	\section{Linear stability of the problem}
	
	By adding a minor perturbation to the equilibrium state with an amplitude of $\epsilon \ll 1$, we use the linear perturbation theory to examine stability. The mathematical representation of this disturbance is as follows:
	
	\begin{widetext}
		\begin{equation}\label{32}
			[\boldsymbol{u},n,\zeta,L,<p>]=[0,n_b,\zeta_b,L_b,<p_b>]+\epsilon [\boldsymbol{u}_1,n_1,\zeta_1,L_1,<\boldsymbol{p}_1>]+O(\epsilon^2)
		\end{equation}
	\end{widetext}
	
	We isolate terms of $O(\epsilon)$ around the equilibrium state by substituting the perturbed variables into equations (\ref{12} to \ref{14}) and linearizing the equations, resulting in the equation shown below
	
	\begin{equation}\label{33}
		\boldsymbol{\nabla}\cdot \boldsymbol{u}_1=0,
	\end{equation}
	
	where $\boldsymbol{u}_1=(U_1,V_1,W_1)$.
	
	The equations (\ref{34}) and (\ref{35}) governing the perturbed variables become
	
	\begin{equation}\label{34}
		Sc^{-1}\left(\frac{\partial \boldsymbol{u_1}}{\partial t}\right)+\sqrt{Ta}(z\times u_1)+\boldsymbol{\nabla} P_{e}+Rn_1\hat{\boldsymbol{z}}=\nabla^{2}\boldsymbol{ u_1},
	\end{equation}
	
	\begin{equation}\label{35}
		\frac{\partial{n_1}}{\partial{t}}+V_c\boldsymbol{\nabla}\cdot(<\boldsymbol{p_b}>n_1+<\boldsymbol{p_1}>n_b)+W_1\frac{dn_b}{dz}=\boldsymbol{\nabla}^2n_1.
	\end{equation}
	
	If we express $G=G_b+\epsilon G_1+O(\epsilon^2)$, where $G=(G_b^c+\epsilon G_1^c)+(G_b^d+\epsilon G_1^d)+O(\epsilon^2)$, then the steady collimated total intensity is perturbed as $L_t\exp\left(\frac{-\kappa\int_z^1(n_b(z')+\epsilon n_1+O(\epsilon^2))dz'}{\cos\theta_0}\right)$. After simplification, we obtain:
	
	\begin{equation}\label{36}
		G_1^c=L_t\exp\left(\frac{-\int_z^1 \kappa n_b(z')dz'}{\cos\theta_0}\right)\left(\frac{\int_1^z\kappa n_1 dz'}{\cos\theta_0}\right).
	\end{equation}
	
	Similarly, $G_1^d$ can be defined as
	
	\begin{equation}\label{37}
		G_1^d=\int_0^{4\pi}L_1^d(\boldsymbol{ x},\boldsymbol{ r})d\Omega.
	\end{equation}
	
	For the radiative heat flux $q=q_b+\epsilon q_1+O(\epsilon^2)$, with $q=(q_b^c+\epsilon q_1^c)+(q_b^d+\epsilon q_1^d)+O(\epsilon^2)$, we find
	
	\begin{equation}\label{38}
		\boldsymbol{q}_1^c=-L_t(\cos\theta_0)\exp\left(\frac{-\int_z^1 \kappa n_b(z')dz'}{\cos\theta_0}\right)\left(\frac{\int_1^z\kappa n_1 dz'}{\cos\theta_0}\right)\hat{z},
	\end{equation}
	
	and
	
	\begin{equation}\label{39}
		q_1^d=\int_0^{4\pi}L_1^d(\boldsymbol{ x},\boldsymbol{r})\boldsymbol{ r}d\Omega.
	\end{equation}
	
	Perturbing the expression for swimming orientation and collecting $O(\epsilon)$ terms yields the perturbed swimming orientation as follows
	
	\begin{equation}\label{40}
		<\boldsymbol{p_1}>=G_1\frac{dM_b}{dG}\hat{\boldsymbol{z}}-M_b\frac{\boldsymbol{q_1}^H}{\boldsymbol{q_b}},
	\end{equation}
	
	where $\boldsymbol{q}_1^H=[\boldsymbol{q}_1^x,\boldsymbol{q}_1^y]$ represents the horizontal component of the perturbed radiative flux $\boldsymbol{q}_1$. By substituting the value of $<\boldsymbol{p_1}>$ from Eq. (\ref{40}) into Eq. (\ref{35}) and simplifying, we obtain
	
	\begin{equation}\label{41}
		\frac{\partial{n_1}}{\partial{t}}+V_c\frac{\partial}{\partial z}\left(M_bn_1+n_bG_1\frac{dM_b}{dG}\right)-V_cn_b\frac{M_b}{q_b}\left(\frac{\partial q_1^x}{\partial x}+\frac{\partial q_1^y}{\partial y}\right)+W_1\frac{dn_b}{dz}=\nabla^2n_1.
	\end{equation}
	
	Eqs. (\ref{34}) and (\ref{35}) can be made simpler by removing both $P_e$ and the horizontal component of $u_1$. This leaves three equations that govern the perturbed variables: the vertical component of velocity ($W_1$), the vertical component of vorticity ($\zeta_1$), and the concentration ($n_1$). The normal modes of these variables can then be further dissected, as shown below:
	
	\begin{equation}\label{42}
		[W_1,\zeta_1,n_1]=[\tilde{W}(z),\tilde{Z}(z),\tilde{N}(z)]\exp{(\sigma t+i(lx+my))}.
	\end{equation}
	
	The equation governing the perturbed intensity $L_1$ can be written as
	
	\begin{equation}\label{43}
		\xi\frac{\partial L_1}{\partial x}+\eta\frac{\partial L_1}{\partial y}+\nu\frac{\partial L_1}{\partial z}+\kappa( n_bL_1+n_1L_b)=\frac{\omega\kappa}{4\pi}(n_bM_1+G_bn_1+A\nu(n_bq_b\cdot\hat{z}-q_bn_1))-\kappa L_bn_1,
	\end{equation}
	
	with the following boundary conditions
	
	\begin{subequations}
		\begin{equation}\label{44a}
			L_1(x, y, z = 1, \xi, \eta, \nu) = 0,\qquad \theta\in[\pi/2,\pi], ~~\phi\in[0, 2\pi],
		\end{equation}
		\begin{equation}\label{44b}
			L_1(x, y, z = 0,\xi, \eta, \nu) = 0,\qquad \theta\in[0,\pi/2], ~~\phi\in[0, 2\pi].
		\end{equation}
	\end{subequations}
	
	The perturbed intensity $L_1^d$ can be represented by the following expression
	
	\begin{equation*}
		L_1^d=\Psi^d(z,\xi,\eta,\nu)\exp{(\sigma t+i(lx+my))}.
	\end{equation*}
	
	From Eqs.~(\ref{36}) and (\ref{37}), we obtain
	
	\begin{equation}\label{45}
		G_1^c=\left[L_t\exp\left(\frac{-\int_z^1 \kappa n_b(z')dz'}{\cos\theta_0}\right)\left(\frac{\int_1^z\kappa n_1 dz'}{\cos\theta_0}\right)\right]\exp{(\sigma t+i(lx+my))}=\mathbb{G}^c(z)\exp{(\sigma t+i(lx+my))},
	\end{equation}
	
	and 
	
	\begin{equation}\label{46}
		G_1^d=\mathbb{G}^d(z)\exp{(\sigma t+i(lx+my))}= \left(\int_0^{4\pi}\Psi^d(z,\xi,\eta,\nu)d\Omega\right)\exp{(\sigma t+i(lx+my))}.
	\end{equation}
	
	Here, $\mathbb{G}(z)$ is the perturbed total intensity, expressed as the sum of $\mathbb{G}^c(z)$ and $\mathbb{G}^d(z)$. Similarly from Eqs.~(\ref{38}) and (\ref{39}), we have
	
	\begin{equation*}
		\boldsymbol{q}_1=[q_1^x,q_1^y,q_1^z]=[P(z),Q(z),S(z)]\exp{[\sigma t+i(lx+my)]},
	\end{equation*}
	
	where
	
	\begin{equation*}
		[P(z), Q(z)]=\int_0^{4\pi}[\xi,\eta]\Psi^d(z,\xi,\eta,\nu) d\Omega,
	\end{equation*}
	
	and
	
	\begin{equation*}
		S(z)=\int_0^{4\pi}[\Psi^c(z,\xi,\eta,\nu)+\Psi^d(z,\xi,\eta,\nu)]\nu d\Omega.
	\end{equation*}
	
	Note that the $P(z)$ and $Q(z)$ appear due to scattering. Conversely, the presence of $S(z)$ is a result of anisotropic scattering, and it diminishes in situations characterized by isotropic scattering.
	
	Now $\Psi^d$ satisfies
	
	\begin{equation}\label{47}
		\frac{d\Psi^d}{dz}+\frac{(i(l\xi+m\eta)+\kappa n_b)}{\nu}\Psi^d=\frac{\omega\kappa}{4\pi\nu}(n_b\mathcal{\mathbb{G}}+G_b\tilde{N}(z)+A\nu(n_b S-q_b\tilde{N}(z)))-\frac{\kappa}{\nu}I_s\tilde{N}(z),
	\end{equation}
	
	subject to the boundary conditions
	
	\begin{subequations}
		\begin{equation}\label{48a}
			\Psi^d( 1, \xi, \eta, \nu) =0,\qquad \theta\in[\pi/2,\pi], ~~\phi\in[0, 2\pi] , 
		\end{equation}
		\begin{equation}\label{48b}
			\Psi^d( 0,\xi, \eta, \nu) =0,\qquad \theta\in[0,\pi/2], ~~\phi\in[0, 2\pi]. 
		\end{equation}
	\end{subequations}
	
	The stability equations reformed as
	
	\begin{equation}\label{49}
		\left(\sigma S_c^{-1}+k^2-\frac{d^2}{dz^2}\right)\left( \frac{d^2}{dz^2}-k^2\right)\tilde{W}=Rk^2\tilde{N}(z),
	\end{equation}
	\begin{equation}\label{50}
		\left(\sigma Sc^{-1}+k^2-\frac{d^2}{dz^2}\right)\tilde{Z}(z)=\sqrt{Ta}\frac{d\tilde{W}}{dz}
	\end{equation}
	\begin{equation}\label{51}
		\left(\sigma+k^2-\frac{d^2}{dz^2}\right)\tilde{N}(z)+V_c\frac{d}{dz}\left(M_b\tilde{N}(z)+n_b\mathbb{G}\frac{dM_b}{dG}\right)-i\frac{V_cn_bM_b}{q_b}(lP+mQ)=-\frac{dn_b}{dz}\tilde{W}(z),
	\end{equation}
	
	with boundary conditions
	
	\begin{equation}\label{52}
		\tilde{W}(z)=\frac{d\tilde{W}(z)}{dz}=\tilde{Z}(z)=\frac{d\tilde{N}(z)}{dz}-V_cM_b\tilde{N}(z)-n_bV_c\mathbb{G}\frac{dM_b}{dG}=0\quad at\quad z=0,
	\end{equation}
	\begin{equation}\label{53}
		\tilde{W}(z)=\frac{d^2\tilde{W}(z)}{dz^2}=\frac{d\tilde{Z}(z)}{dz}=\frac{d\tilde{N}(z)}{dz}-V_cM_b\tilde{N}(z)-n_bV_c\mathbb{G}\frac{dM_b}{dG}=0\quad at\quad z=1.
	\end{equation}
	
	in which the non-dimensional wavenumber $k$ is defined as the square root of the sum of the squares of $l$ and $m$. Equations (\ref{49})-(\ref{51}) collectively constitute an eigenvalue problem, in which $\sigma$ is expressed as a function dependent on various dimensionless parameters, including $V_c$, $\kappa$, $\omega$, $A$, $l$, $m$, $\theta_i$, and $R$.
	
	An alternative representation of Equation (\ref{51}) is as follows
	
	\begin{equation}\label{54}
		D^2\tilde{N}(z)-\Lambda_3(z)D\tilde{N}(z)-(\sigma+k^2+\Lambda_2(z))\tilde{N}(z)-\Lambda_1(z)\int_1^z\tilde{N}(z) dz-\Lambda_0(z)=Dn_b\tilde{W}, 
	\end{equation}
	
	where
	
	\begin{subequations}
		\begin{equation}\label{55a}
			\Lambda_0(z)=V_cD\left(n_b\mathbb{G}^d\frac{dM_b}{dG}\right)-i\frac{V_cn_bM_b}{q_b}(lP+mQ),
		\end{equation}
		\begin{equation}\label{55b}
			\Lambda(z)=\left(\frac{\kappa}{\cos\theta_0}\right) V_cD\left(n_bG_b^c\frac{dM_b}{dG}\right)
		\end{equation}
		\begin{equation}\label{55c}
			\Lambda_2(z)=2\left(\frac{\kappa}{\cos\theta_0}\right) V_c n_b G_b^c\frac{dM_b}{dG}+V_c\frac{dM_b}{dM}DG_b^d,
		\end{equation}
		\begin{equation}\label{55d}
			\Lambda_3(z)=V_cM_b.
		\end{equation}
	\end{subequations}
	
	Introducing a new variable denoted as
	
	\begin{equation}\label{56}
		\tilde{\Theta}(z)=\int_1^z\tilde{N}(z')dz',
	\end{equation}
	
	the linear stability equations can be reformulated as
	
	\begin{equation}\label{57}
		D^4\tilde{W}-\left(\sigma S_c^{-1}+k^2\right)D^2\tilde{W}-\left(\sigma S_c^{-1}+k^2\right)\tilde{W}=Rk^2D\tilde{\Theta},
	\end{equation}
	
	\begin{equation}\label{58}
		D^2\tilde{Z}(z)-\left(\sigma S_c^{-1}+k^2-D^2\right)\tilde{Z}(z)=\sqrt{Ta}D\tilde{W},
	\end{equation}
	
	\begin{equation}\label{59}
		D^3\tilde{\Theta}-\Lambda_3(z)D^2\tilde{\Theta}-(\sigma+k^2+\Lambda_2(z))D\tilde{\Theta}-\Lambda_1(z)\tilde{\Theta}-\Lambda_0(z)=Dn_b\tilde{W}.
	\end{equation}
	
	The boundary conditions remain the same except for the flux boundary condition, which is given by the equation
	
	\begin{equation}\label{60}
		D^2\tilde{N}(z)-\Lambda_2D\tilde{N}(z)-n_bV_c\mathbb{G}\frac{dM_b}{dG}=0\quad \text{at}\quad z=0,1.
	\end{equation}
	
	Furthermore, an additional boundary condition is introduced
	
	\begin{equation}\label{61}
		\tilde{\Theta}(z)=0,\quad \text{at}\quad z=1.
	\end{equation}
	
	
	\section{NUMERICAL RESULTS}
	The system of equations specifically denoted as Eqs.~(\ref{57})-(\ref{59}), is solved using the iterative Newton-Raphson Kantorovich (NRK) method. To identify the most unstable mode, we initiate with an initial equilibrium solution and employ a specific set of constant parameters. These parameters include $Sc = 20$, $G_c = 1$, and $V_c$ with values of $10$, $15$, and $20$. We vary the scattering albedo $\omega$ within the range of 0 to 1. Additionally, we consider two values for $\kappa$, namely, 0.5 and 1, and explore different values for $\theta_i$, specifically, 0, 40, and 80. Furthermore, we investigate the influence of varying the Taylor number, which represents the rotation rate, across a range of values from zero to higher values.
	
	In our examination of the impact of rotation on forward scattering in an algal suspension, we focus on specific values of the scattering albedo $\omega$. We select the value of $\omega$ in such a manner that the sublayer's position coincides with the midpoint of the domain for collimated particle flux (normal illumination, i.e., $\theta_i=0$). As the angle of irradiation, $\theta_i$, increases, the sublayer's location shifts towards the upper region of the suspension. Here, we consider three values of $\theta_i$, specifically 0, 40, and 80, for which the sublayer's location aligns with the midpoint, three-quarters, and the uppermost part of the suspension, respectively.
	
	\subsection{WHEN SCATTERING IS WEAK}
	
	This study's main goal is to find out how rotation influences the start of bioconvection. In order to better understand this phenomenon, researchers are comparing the effects of two variables: self-shading and scattering. They do this by utilising an albedo value for scattering that is smaller, represented as $\omega$. Additionally, by altering the extinction coefficient $\kappa$, they may change how much self-shading occurs. A larger value of $\kappa=1$ denotes a substantial self-shading impact, whereas a lower value of $\kappa=0.5$ denotes a weaker self-shading influence.
	
	
	\subsubsection{$V_c=15$}
	(\romannumeral 1) $\kappa=0.5$
	
	In this section, we explore how variations in the Taylor number influence the occurrence of bioconvective instability across three specific values of the forward scattering coefficient, designated as "A" and at three different values of $\theta_i$. This examination is conducted while maintaining a set of constant parameters: $V_c=15$, $\kappa=0.5$, and $\omega=0.46$.
	
	\begin{figure*}[!htbp]
		\includegraphics{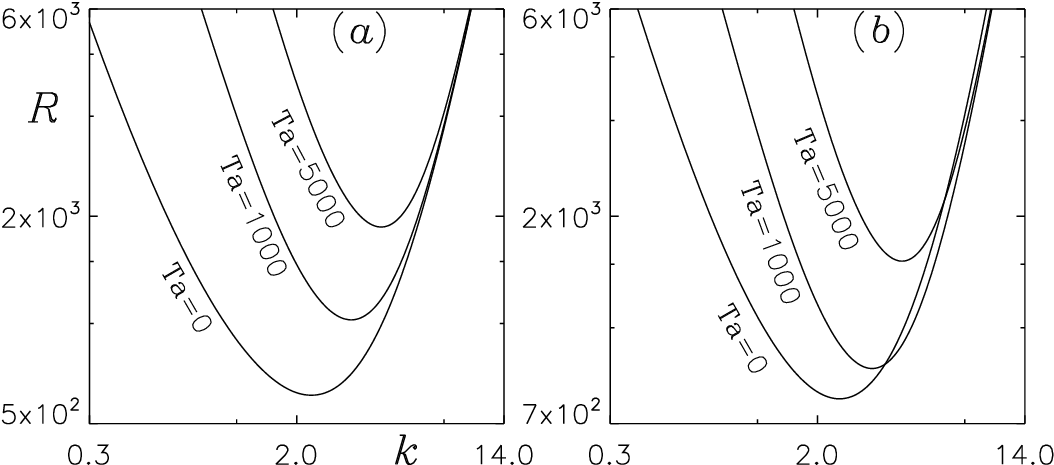}
		\caption{\label{fig3} The marginal stability curves for (a) $A=0$, (b) $A=0.4$. Here, the other governing parameter values $Sc=20,V_c=15,k=0.5$, $\omega=0.46$ and $\theta_i=0$ are kept fixed.}
	\end{figure*}
	
	At $\theta_i=0$ and $A=0$, the sublayer is positioned at approximately $z=0.6$ of the domain's height. As the parameter $A$ increases, the sublayer's location shifts toward the middle of the domain. When $A=0$, the Taylor number ($Ta$) exhibits variation up to a nonzero threshold. At $Ta=0$, the linear stability analysis predicts a stationary perturbation mode in the basic equilibrium state. As $Ta$ rises to 5000, the solutions' behaviour remains consistent, but the critical Rayleigh number increases while the pattern wavelength decreases (refer to Fig. \ref{fig3}(a)). For values of $A$ at 0.4 and 0.8, the perturbation mode remains unaltered (see Fig.~\ref{fig3}).
	
	\begin{figure*}[!bt]
		\includegraphics{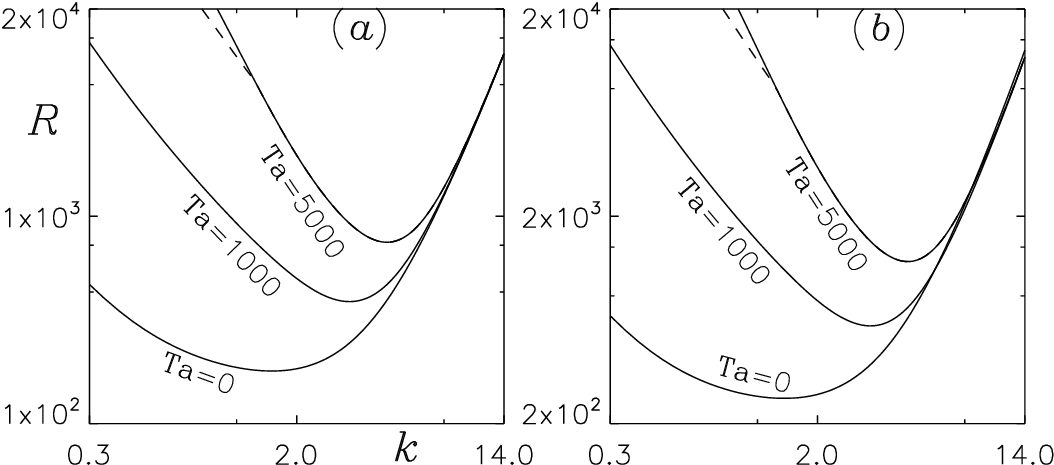}
		\caption{\label{fig4} The marginal stability curves for (a) $A=0$, (b) $A=0.4$. Here, the other governing parameter values $Sc=20,V_c=15,k=0.5$, $\omega=0.46$ and $\theta_i=40$ are kept fixed.}
	\end{figure*}	
	
	When $\theta_i=40$, the sublayer in the equilibrium state is located at the three-quarter height of the suspension for $A=0$ and as $A$ increases the location of the sublayer at the equilibrium state shifts towards the mid-height of the domain. At $Ta=0$, a stationary mode of the bioconvective solution is observed for $A=0$. As $Ta$ is increased to 1000, the most unstable solution still occurs on the stationary branch of the neutral curve, resulting in a When $\theta_i=40$, the sublayer in the equilibrium state is positioned at three-quarters of the suspension's height for $A=0$, and as $A$ increases, it shifts towards the middle of the domain. At $Ta=0$, a stationary bioconvective solution is observed for $A=0$. When $Ta$ is increased to 1000, the most unstable solution still lies on the stationary branch of the neutral curve, resulting in a stationary solution. However, at $Ta=5000$, an oscillatory branch emerges from the stationary branch of the marginal stability curve at approximately $k=1.82$, persisting for all $k<1.82$. Nonetheless, the most unstable bioconvective solution remains on the stationary branch, resulting in a stationary solution. Similar behaviour is exhibited for $A=0.4$ and $A=0.8$. In these cases, the critical Rayleigh number increases, and the pattern wavelength decreases as the Taylor number increases for all subcases (see Fig.~\ref{fig4}). solution. But, at $Ta=5000$, an oscillatory branch bifurcates from the stationary branch of the marginal stability curve at $k\approx1.82$ and exists for all $k<1.82$. However, the most unstable bioconvective solution still occurs on the stationary branch, resulting in a stationary solution. For $A=0.4$ and 0.8, the bioconvective solutions show similar behaviour. In this case, the critical Rayleigh number increased and pattern wavelength decreased as the Taylor number increased for all subcases (see Fig.~\ref{fig4}).
	
	\begin{figure*}[!htbp]
		\includegraphics{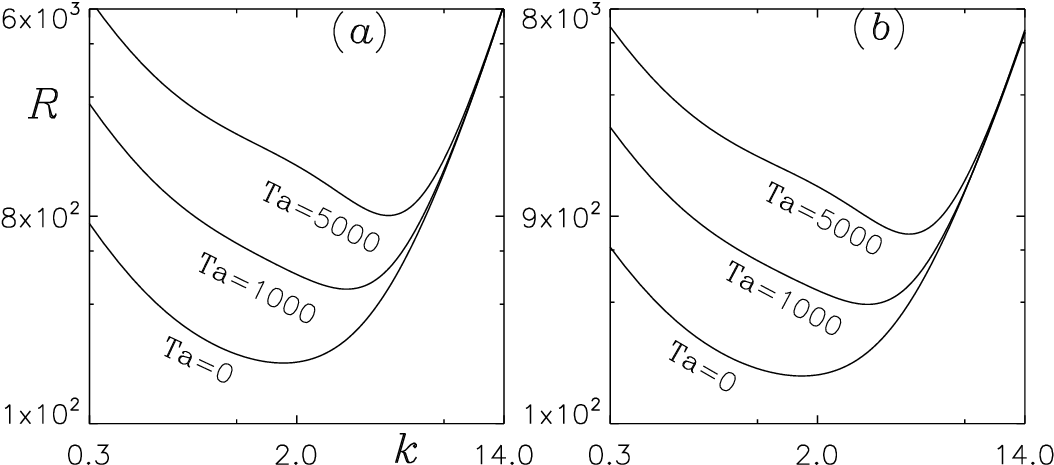}
		\caption{\label{fig5} The marginal stability curves for (a) $A=0$, (b) $A=0.4$. Here, the other governing parameter values $Sc=20,V_c=15,k=0.5$, $\omega=0.46$ and $\theta_i=80$ are kept fixed.}
	\end{figure*}
	
	When $\theta_i=80$, the sublayer is situated near the top of the domain when $A=0$, and as $A$ increases, it shifts towards three-quarters of the domain's height. For $A=0$, the Taylor number ($Ta$) varies from 0 to a nonzero threshold. At $Ta=0$, the linear stability analysis predicts a stationary perturbation mode in the basic equilibrium state. As $Ta$ increases up to 5000, the solutions' behaviour remains consistent, but the critical Rayleigh number increases while the pattern wavelength decreases (see Fig.\ref{fig5}(a)). For $A=0.4$ and $A=0.8$, the perturbation mode remains unchanged (refer to Fig.\ref{fig5}).

	(\romannumeral 2) $\kappa=1$
	
	In this section, we investigate how the Taylor number impacts bioconvective instability across three distinct values of the forward scattering coefficient, denoted as $A$, and at three different values of $\theta_i$. This analysis is carried out while maintaining a set of constant parameters, specifically $V_c=15$, $\kappa=1$, and $\omega=0.59$.
	
	\begin{figure*}[!htbp]
		\includegraphics{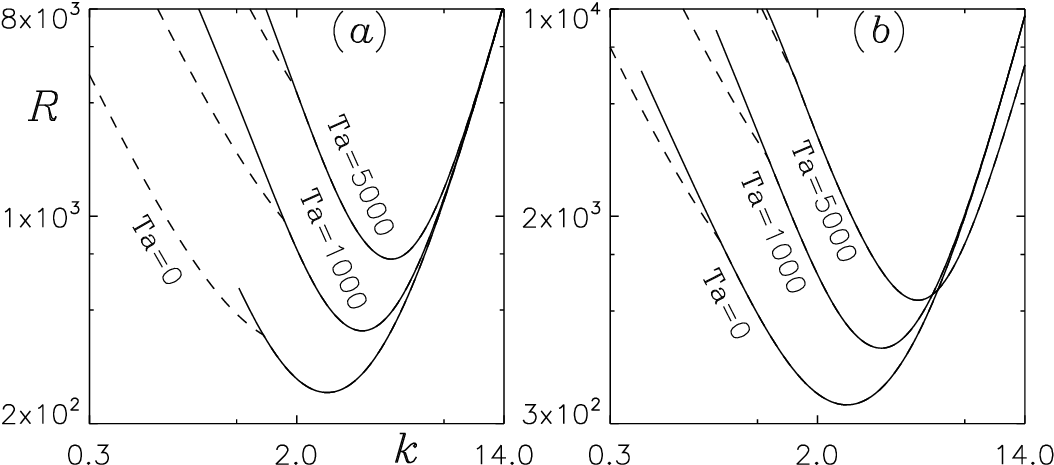}
		\caption{\label{fig6} The marginal stability curves for (a) $A=0$, (b) $A=0.4$. Here, the other governing parameter values $Sc=20,V_c=15,k=1$, $\omega=0.59$ and $\theta_i=0$ are kept fixed.}
	\end{figure*} 
	
	For $\theta_i=0$, when $A=0$, the sublayer is located approximately at $z=0.62$ within the suspension. As the value of $A$ increases, the sublayer at the equilibrium state shifts towards the middle of the domain. In this scenario, the Taylor number ($Ta$) is varied from 0 to a higher value of $Ta=5000$. At $Ta=0$, an oscillatory branch emerges from the stationary branch of the marginal stability curve at $k\approx1.92$, and this oscillatory branch exists for all $k<1.92$. Notably, the most unstable solution occurs on the oscillatory branch of the marginal stability curve, resulting in an overstable solution. When $Ta$ is increased to 1000, a similar oscillatory branch splits from the stationary branch, but the most unstable solution now occurs on the stationary branch of the neutral curve, leading to a stationary solution and an increase in the critical Rayleigh number. For $Ta=5000$, the same nature of the marginal stability curve and the bioconvective solution is observed, and the critical Rayleigh number continues to increase (see Fig.\ref{fig6}(a)). This behaviour remains consistent for $A=0.4$ and $A=0.8$, as shown in Fig.\ref{fig6}.
	
	\begin{figure*}[!htbp]
		\includegraphics{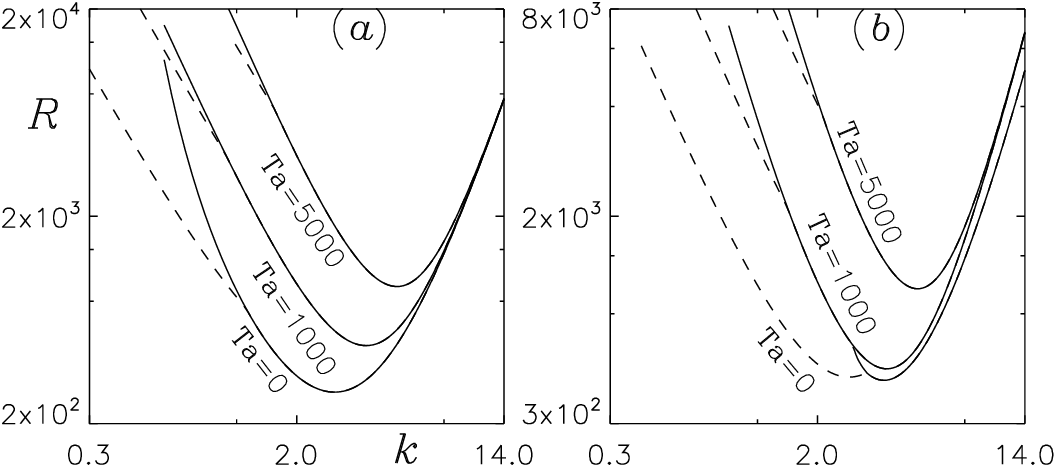}
		\caption{\label{fig7} The marginal stability curves for (a) $A=0$, (b) $A=0.4$. Here, the other governing parameter values $Sc=20,V_c=15,k=1$, $\omega=0.59$ and $\theta_i=40$ are kept fixed.}
	\end{figure*}
	
	For $\theta_i=40$, the sublayer in the equilibrium state is positioned at three-quarters of the suspension's height when $A=0$. At $Ta=0$, an oscillatory branch branches off from the stationary branch of the marginal stability curve at $k\approx1.7$, and this oscillatory branch is present for all $k<1.7$. Similar to the previous case, the most unstable solution is found on the oscillatory branch of the marginal stability curve, resulting in an overstable solution. As $Ta$ increases to 1000, another oscillatory branch forms from the stationary branch, but the most unstable solution now resides on the stationary branch of the neutral curve, leading to a stationary solution and an increase in the critical Rayleigh number. For $Ta=5000$, the same characteristics of the marginal stability curve and bioconvective solution persist, with the critical Rayleigh number continuing to rise (see Fig.\ref{fig7}(a)). This behaviour is also consistent for $A=0.4$ and $A=0.8$, as depicted in Fig.\ref{fig7}.
	
	\begin{figure*}[!htbp]
		\includegraphics{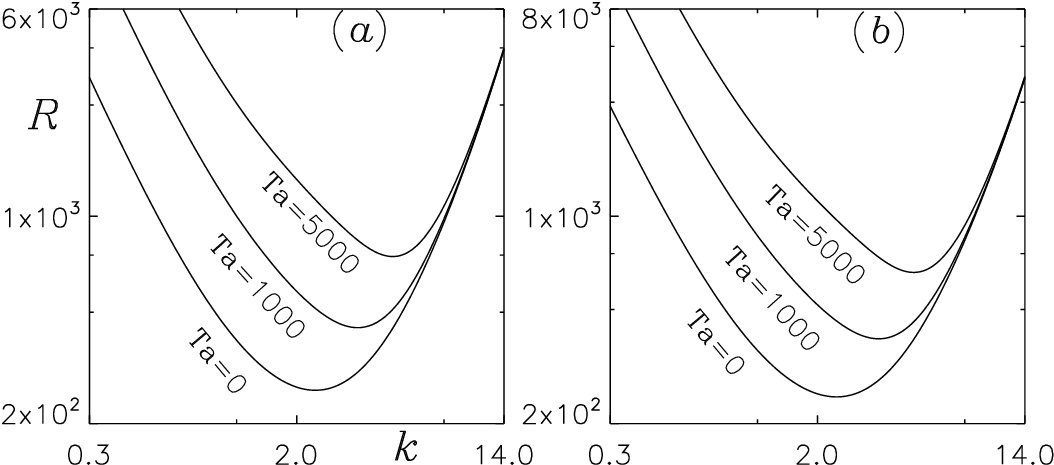}
		\caption{\label{fig8} The marginal stability curves for (a) $A=0$, (b) $A=0.4$. Here, the other governing parameter values $Sc=20,V_c=15,k=1$, $\omega=0.59$ and $\theta_i=80$ are kept fixed.}
	\end{figure*}

	For $\theta_i=80$, the sublayer in the equilibrium state is located at the top of the suspension when $A=0$. At $Ta=0$, the most unstable solution is found on the stationary branch of the marginal stability curve, resulting in a stable solution. As $Ta$ increases to 1000", the same characteristics of the marginal stability curve and bioconvective solution are observed, and the critical Rayleigh number continues to increase (see Fig.\ref{fig8}(a)). When the value of $Ta$ further increases to 5000", the most unstable solution is again on the stationary branch of the neutral curve, resulting in a stationary solution (see Fig.\ref{fig8}). The behaviour of the perturbation remains similar for $A=0.4$ and $A=0.8$. In all these cases, a common phenomenon is observed: the critical Rayleigh number increases and the pattern wavelength decreases as the Taylor number (rotation rate) increases.

	\section{Conclusion}
	
	The proposed phototaxis model aims to explore how rotation influences the onset of light-induced bioconvection in a forward-scattering algal suspension exposed to oblique collimated light. This model takes into account a linear anisotropic scattering coefficient.
	
	Through linear stability analysis, it is predicted that as the Taylor number (representing the rotation rate) increases, stationary disturbances transform into oscillatory disturbances. These oscillatory bioconvection solutions are often observed at higher values of cell swimming speed and absorption with a lower angle of incidence. Notably, the pattern wavelength decreases while the corresponding Rayleigh number increases with an increasing Taylor number. This suggests that the suspension becomes more stable as the rotation rate rises.
	
	Compared to the model presented by Rajput, the proposed model is considered more realistic because it accounts for the anisotropic scattering of light by the algae cells. In contrast, Rajput's model assumed isotropic scattering by the algae cells, which might not be accurate due to the actual shape of these cells. However, it is important to acknowledge that the suggested phototaxis model should be validated by comparing its theoretical hypotheses with quantitative experimental data on bioconvection in a purely phototactic algal suspension. Regrettably, such data are not currently available.
	
	Hence, there is a need to identify suitable microorganism species primarily exhibiting phototactic behaviour, as most naturally occurring algal species tend to be gravitactic or gyrotactic with some phototactic tendencies. It's worth emphasizing that the suggested model has the potential to simulate naturally occurring phenomena once validated through empirical research.
	
	\section*{Data Availability}
	
	The article includes the necessary information and analysis to support the conclusions and results reported.
	
	\nocite{*}
	
	
	\bibliography{aniso_oblique_rotation}
	
\end{document}